# Spin-statistics-quantum number connection and supersymmetry

Richard M. Weiner[i]

Laboratoire de Physique Théorique, Univ. Paris-Sud, Orsay, France and

Physics Department, University of Marburg, Germany

PACS: 11.30.Fs, 11.30.Pb, 11.30.Rd, 12.39.De, 14.80.Ec

The analogy between the Skyrme and Higgs field leads to the conjecture that all fermions are skyrmions and thus always carry conserved quantum numbers, which are identified with baryon or lepton quantum numbers. This connection between spin and quantum numbers, which parallels the connection between spin and statistics due to the Pauli principle, may explain why supersymmetry has not been observed. Creation of s-particles at higher than present energies due to a breakdown of the Skyrme mechanism might imply the violation of the exclusion principle.

Fermions, i.e. half odd integer spin particles, are characterized by two specific properties which distinguish them from integer spin particles (bosons): they obey the Pauli exclusion principle and possess conserved baryon or lepton quantum numbers. The first property makes part of the fundamental principles of quantum mechanics and has wide applications in physics and chemistry, constituting the basis of the spin-statistics theorem. It was first discovered by Pauli in atomic physics and preceded by decades the discovery of the quantum number property, which had to await the successive discovery of elementary particles. Both properties are of experimental nature and have resisted so far not only attempts of theoretical derivation but even of intuitive explanation. For the Pauli principle this state of affairs is resumed e.g. in a 500 pages collection of articles by Duck and Sudarshan [1]. The issue of the non-understanding of the exclusive association of baryon or lepton quantum numbers with fermions is tantamount to the ignorance of the mechanism of supersymmetry breaking [2].

---

[i] Electronic address: weiner@staff.uni-marburg.de

However, while the Pauli principle is usually assumed to be a fundamental, energy independent property of fermions, the quantum number property has been assumed by many authors as specific for low energies, the more so that its negation in the form of supersymmetry possibly leads to solutions of several other puzzles of high energy physics. Indeed, supersymmetry (SUSY) which claims that for every type of fermion there exists a corresponding type of boson with the same baryon or lepton quantum number and the same mass, and vice-versa, has been considered a popular cure for several unsolved theoretical problems of particle physics [2]. These include a) gauge coupling unification, b) dark matter and the c) hierarchy problem. Furthermore d) SUSY is a fundamental postulate of string theory.

Although these considerations have been viewed by many as arguments for the expectation that SUSY is a symmetry of nature, one must not forget that there are theoretical arguments which plea against that. SUSY is a strong postulate, since it establishes a symmetry between classical and quantum physics, which up to now has not been observed in any other domain. While any number of bosons can occupy the same quantum state, for fermions this is not possible, because of the exclusion principle, which allows only one fermion in a given state. But when the occupation numbers become large quantum physics approaches the classical limit. This means that while bosons also exist in classical physics, fermions do not. That makes it difficult to expect that bosons, *if at all*, possess the same quantum numbers as fermions. (This remark refers to baryon or lepton quantum numbers, which are coupled to short range fields, and not to electric, long range charges, cf. also below).

Another argument which argues against SUSY is related to its breaking. While the breaking of a symmetry itself is not necessarily a drawback - most symmetries which have been observed in nature are broken - the breaking of superymmetry, reflected among other things in the differences of masses between particles and their supersymmetric s- partners, is, as mentioned above, not understood and this makes SUSY unfalsifiable. Indeed, up to now no superymmetric partners of presently known particles have been observed. (As a matter of fact, the long awaited recent experimental results at the CERN LHC collider seem to rule out supersymmetry at least in its simplest version [3]). This means that if SUSY is really a symmetry of nature, the masses of s-particles probably exceed the presently available energy limits. However the minimum s-mass is unknown, which means that in principle it can be pushed upwards to arbitrarily large values, without disproving supersymmetry. Actually, the difference between classical and quantum physics could suggest that this minimum mass is of the order of the Planck mass $M_{Planck} = \sqrt{(\hbar c / G_N)}$, where $G_N$ is the gravitational constant, since at energies of this order the difference between classical and quantum physics might disappear, everything becoming quantized, possibly also space-time, implying space-time non-commutativity [4].

In the following I present a concrete physical model according to which fermions are skyrmions, i.e. topological solitons, which is consistent with the standard model and which explains why supersymmetry has not been observed so far. In this model all fermions and only fermions have baryon or lepton quantum numbers and these are of topological nature. In this way the fact that these quantum numbers have been observed only for fermions gets a natural explanation and is reflected in a an extension of the spin-statistics connection to the spin-statistics-quantum number connection.

That, in contradiction to SUSY, fermions are *always* associated with conserved quantum numbers was suggested by Finkelstein and Rubinstein [5] in terms of a non-linear field theory that admits kinks, i.e. "extended, indestructible objects whose number is conserved". The Finkelstein-Rubinstein model (FRM) preceded not only supersymmetry but also the standard model and since the symmetries associated with the baryon or lepton quantum numbers have been considered in the standard model as *accidental*, i.e. effects to be explained by a theory which supersedes the standard model, the possible link between FRM and the standard model did not get much attention in the literature.

On the other hand the Finkelstein-Rubinstein model was inspired by the Skyrme [6] approach to strong interactions in which baryons are considered as topological solitons, and this approach [7] got in the 1980-s new support, among other things, from the phenomenological success of the non-linear sigma model to which it reduces in a limit, from the fact that it could account qualitatively for certain static properties of baryons [8] including their mass and from the link between the Skyrme model and QCD [9], [10].

Skyrme [6] suggested an explanation why spin ½ nucleons have baryon quantum numbers by postulating the existence of a quaternion-valued scalar field

$$U_{Skyrme} = \exp[(2i/f) \, \boldsymbol{\tau} \cdot \boldsymbol{\pi}] \qquad (1)$$

the Lagrangian of which is non-linear:

$$\mathcal{L}_{Skyrme} = (1/16) \, f^2 \, \mathrm{Tr} \, (\partial_\mu U_{Skyrme} \partial^\mu U_{Skyrme}^\dagger) +$$
$$+ (1/32e^2) \, \mathrm{Tr} \, [(\partial_\mu U_{Skyrme}) \, U_{Skyrme}^\dagger, (\partial_\nu U_{Skyrme}) \, U_{Skyrme}^\dagger]^2. \qquad (2)$$

Here f is the pion decay constant, e a dimensionless parameter, $\boldsymbol{\pi}$ the pseudo-scalar pion triplet field and $\boldsymbol{\tau}$ represents the isospin Pauli matrices. Skyrme proved that under certain conditions such a scalar field can lead to topological solitons, i. e. static classical solutions of finite size and energy and with a conserved quantum number, which may have half odd integer spin and which can be identified with the baryon number.

Two decades later, using the $N_{colour}$ expansion due to t'Hooft [11], Witten [10] proved that for odd $N_{colour}$ skyrmions actually have half odd spin and that ordinary baryons can be understood as solitons in current algebra effective Lagrangians. While the identification by Skyrme of baryons with the soliton solution of Eqs. (1), (2), was based on pure phenomenological arguments, that of Witten, although also only of qualitative nature [12], appears more fundamental, given its link with QCD. More concretely, while Skyrme proved that the solutions of (1), (2) have both half odd integer and integer spin and gave phenomenological arguments for identifying the first one with the nucleon, Witten proved that a Lagrangian of type (1) completed with a Wess-Zumino term, which takes care of the chiral anomaly, leads only for odd values of $N_{colour}$, and in particular for $N_{colour} = 3$, as in QCD, to half odd spin.

New evidence for the soliton character of the baryon may have been found with the possible discovery [13] of excited baryons made of five quarks, which had been predicted by the QCD version of the sigma model.

If, as described above, a scalar field of Skyrme type may explain the connection between spin and baryon quantum number of the nucleon, the question arises whether a similar mechanism does not apply to the analogous problem of spin and leptonic quantum number. Actually, as will be argued below, the above question is even more justified than for baryons. This question has been recently [14] answered in the positive: Using the analogy between the scalar Skyrme field and the scalar Higgs field, it was suggested that all fermions are topological solitons. Given the relevance of this analogy for the present considerations we will review its essential points below.

According to the standard model in the electroweak sector there also exists a non-linearly interacting scalar field, the Higgs field, which formally satisfies an equation similar to (1) (cf. below). In analogy to the Skyrme field, which accounts for the spontaneously broken $SU(2)_L$ x $SU(2)_R$ symmetry of the non-linear sigma model in strong interactions, the Higgs field breaks spontaneously the electroweak SU(2) x U(1) symmetry. And last but not least, while the scalar Skyrme field is only a theoretical construction, which can be tested only by its indirect consequences, the recent LHC discovery of a Higgs-like particle provides strong evidence that the scalar Higgs field is a real field. This strengthens the argument that leptons, too, are solitons.

As a matter of fact, the similarity between the Higgs and the Skyrme field was observed a long time ago. It has been known that if the Higgs mass exceeds a certain limit of the order of $1/\sqrt{G_F} \approx 300$ GeV, where $G_F$ is the Fermi constant, at energies of this order weak interactions become strong and the SU(2) x U(1) electroweak sector of the standard model could also be approximated by a gauged non-linear sigma model [15]. Based on this result, Gipson and Tze [16] found that the electroweak sector, too, admits topological solitons, which could behave, what concerns spin and quantum numbers, as leptons. This observation, though, cannot be used straightforwardly in our problem, because in the Gipson-Tze approach leptons are very heavy and interact strongly. However, as shown by Tie-zhong Li [17], if one considers the more general case of a Higgs field represented by two doublets rather than one, as is the case in the minimal standard model, the analogy between the Skyrme and the Higgs field applies also for light and weakly interacting leptons. Indeed, consider the generalized expression of the Skyrme Lagrangian [18], which holds also for non-vanishing pion masses $m_\pi$, when the chiral symmetry is broken

$$\mathcal{L}_{Skyrme} = (1/16)\, f^2\, Tr\, (\partial_\mu U_{Skyrme} \partial^\mu U_{Skyrme}^\dagger) +$$
$$+ (1/32e^2)\, Tr\, [(\partial_\mu U_{Skyrme})\, U_{Skyrme}^\dagger, (\partial_\nu U_{Skyrme})\, U_{Skyrme}^\dagger]^2 +$$
$$+ (1/8)\, m_\pi^2\, f^2\, (Tr\, U_{Skyrme} - 2)\,. \qquad (3)$$

Using (1) Eq. (3) can be written as an expansion in terms of the pion field $\boldsymbol{\pi}$

$$\mathcal{L}_{Skyrme} = (1/2)\, \partial_\mu \boldsymbol{\pi} \cdot \partial^\mu \boldsymbol{\pi} - (1/2)\, m_\pi^2\, \boldsymbol{\pi} \cdot \boldsymbol{\pi} + O(\pi^4) \qquad (4)$$

On the other hand, if the Higgs consists of two doublets, then among the surviving five physical Higgs particles there exists a pseudo-scalar triplet **H**, the Lagrangian of which reads [17]

$$\mathcal{L}_{Higgs} = (1/2)\, \partial_\mu \mathbf{H} \cdot \partial^\mu \mathbf{H} - (1/2)\, m_H^2\, \mathbf{H} \cdot \mathbf{H} + O(\mathbf{H}^4) \qquad (5)$$

where $m_H$ is the Higgs mass. Expressing **H** in terms of a quaternion field

$$U_{Higgs} = \exp[(2i/F)\, \boldsymbol{\tau} \cdot \mathbf{H}] \qquad (6)$$

the formal similarity between Eqs. (1) and (6) and Eqs. (4) and (5) respectively shows that the Higgs Lagrangian can be written as

$$\mathcal{L}_{Higgs} = (1/16)\, F^2\, Tr\, (\partial_\mu U_{Higgs}\, \partial^\mu U_{Higgs}^\dagger) +$$
$$+ (1/32E^2)\, Tr\, [(\partial_\mu U_{Higgs})\, U_{Higgs}^\dagger, (\partial_\nu U_{Higgs})\, U_{Higgs}^\dagger]^2$$
$$+ (1/8)\, m_H^2 F^2 (Tr\, U_{Higgs} - 2)\,. \qquad (7)$$

Here F and E are constants analogous to f and e in Eqs. (2, 3).

In the limit of small masses $m_H$ the last term in Eq. (7) can be neglected and we are left with

$$\mathcal{L}_{Higgs} = (1/16) \, F^2 \, \text{Tr} \, (\partial_\mu U_{Higgs} \, \partial^\mu U_{Higgs}{}^\dagger) +$$

$$+ (1/32E^2) \, \text{Tr} \, [(\partial_\mu U_{Higgs}) \, U_{Higgs}{}^\dagger, (\partial_\nu U_{Higgs}) \, U_{Higgs}{}^\dagger]^2 \quad (8)$$

which is the equivalent of Eq.(2).

We conclude that the analogy between the Skyrme Lagrangian and the Higgs Lagrangian holds also for small Higgs masses [19]. Since for small values of $m_H$ the Higgs field is weakly interacting and since the formation of a soliton does not depend on the strength of the interaction, the Skyrme approach would apply also to weakly interacting leptons.

Furthermore, the lepton-quark symmetry apparently present in the standard model [20] suggests the extension of this approach to quarks leading to the conjectures:

(a) *all known fermions are topological solitons.*

This postulate implies that the Skyrme mechanism may be responsible for the observation that all known fermions have baryon or lepton quantum numbers. Moreover the absence of bosonic leptons and baryons [21] proves that at least at present energies

*(b) this is the only mechanism nature chooses to produce baryon and lepton quantum numbers.*

Conversely, if at energies higher than a critical energy the Skyrme mechanism ceases to be the only mechanism through which nature creates fermions, supersymmetry might be established and s-particles created.

An important by-product of the topological origin of the baryonic and leptonic quantum numbers is the natural and welcome feature that in contrast to electric type charges the topological charges are not coupled to long-range fields and are measured through simple counting. Formally their law of combination is the same as that of the additivity of the homotopy classes in the homotopy group $\pi_3\,(S^3) \approx Z_\infty$, the additive group of the integers. Topological charges depend only on the element of the homotopy group and not on the behaviour of the field at finite space-time coordinates or on the form of the Lagrangian. This distinguishes them from Noether charges and supports conjecture b), i.e. that baryonic and leptonic quantum numbers are found only with fermions, while Noether charges are associated both with fermions and bosons. In this way the Skyrme-Higgs meachanism turns out to be responsible not only for the masses of fermions but also for their quantum numbers.

Another important consequence of the assertion that solitons are the source of baryon and lepton quantum number and half odd spin is that there should not exist massless leptons and quarks, because solitons defined by Eqs. (2), (8) are stable and have a non-vanishing minimum energy [7]. Here from would follow that 1) *only particles without baryon and lepton quantum numbers can be massless;* 2) *even the lightest neutrino has non-vanishing mass* [22]. Finally it was mentioned in the introduction that supersymmetry might be restored at energies of the order of the Planck mass. However, as will be argued below, there are reasons to believe that supersymmetry is associated with the breakdown of the Pauli principle. In this case the restoration of supersymmetry even at Planck energies seems to be ruled out by the Chandrasekhar limit of the masses of white dwarfs. Indeed the value of this limit of 1.4 solar masses, which is in agreement with experimental data was obtained [23] by Chandrasekhar assuming that in these stars the gravitational attraction energy overcomes the repulsive Pauli energy of the degenerate electron gas and the stars collapse to black holes. Moreover, the particular value of the Chandrasekhar limit follows if one assumes that electrons satisfy the ultra-relativistic limit of Fermi-Dirac statistics $\varepsilon \sim n^{4/3}$, while the non-relativistic limit of the energy density $\varepsilon \sim n^{5/3}$ considered by Chandrasekhar's predecessors for the stellar electron gas leads to higher mass limits for white dwarfs, in contradiction with experiment. This suggests that the Pauli principle reflected in the Fermi-Dirac statistics holds up to Planck energies.

The preceding considerations justify the question whether SUSY is at all a symmetry of nature. Besides that and more importantly, the consequences of the fundamental statement that all fermions are topological solitons, which is at the core of the above considerations, can easily be falsified by the discovery of a single leptonic or baryonic boson. As pointed out in the introduction, supersymmetry is not in such a favourable position and cannot be falsified. And last but not least, so far there is no experimental evidence for supersymmetry, to be opposed by the compelling counterevidence for it, the persistent absence of bosonic leptons and baryons.

Independent of the identification of fermions with topological solitons, the fact that all fermions possess lepton and baryon quantum numbers is an experimental fact which is, in the present context, as well established as the connection between fermions and the Pauli principle, which determines the Fermi-Dirac statistics. This leads unavoidably, at least for present energies, to an *extension* of the spin-statistics connection, to the *spin-statistics-quantum numbers connection,* which, in this larger formulation, should be stated as follows: half odd spin particles obey Fermi-Dirac statistics <u>and</u>

<u>possess baryon or lepton quantum numbers</u>; integral spin particles obey Bose-Einstein statistics <u>and do not possess baryon or lepton quantum numbers</u>. The words underlined are missing in the old formulation. Furthermore the fact that the Pauli principle goes hand in hand with the exclusive association of lepton and baryon quantum numbers with half odd spin suggests a common mechanism responsible for these two observations and in particular that a possible violation of this exclusive association represented by supersymmetry may be accompanied by a violation of the Pauli principle. Actually in ref. [14] independent arguments were presented which suggest that the exclusion principle is also a consequence of the solitonic nature of fermions. This means that a possible observation of s-particles at higher than present energies, as a consequence of the breakdown of the Skyrme mechanism, might be associated with a violation of the Pauli principle. This could open quite a new avenue in the search for supersymmetry and physics beyond the standard model. So far, however, even when in a QFT such as the one on a noncommutative space-time the Lorentz invariance is violated, the Pauli spin-statistics relation still remains unbroken [24]. For a comprehensive discussion of this issue cf. Ref. [25]. Besides that, the breakdown of the Pauli principle at energies above the Higgs mass might have dramatic consequences for the usefulness of SUSY in solving the hierarchy problem, which consists in protecting the Higgs mass from huge renormalization effects. Indeed, in the presence of supersymmetry this protection is supposed to be realized by the cancellation, due to the Pauli principle, of the quadratic mass renormalization effects of the fermionic loops by the renormalization effects of the loops of the corresponding bosonic s-partners. In the absence of the Pauli principle this cancellation, obviously, does not occur.

The analogy between the Skyrme and Higgs field, which is reflected in the assumption that leptons are also topological solitons, suggests that leptons too are composite objects and that the electroweak sector of the standard model is in a certain sense in a period of evolution reminiscent of the status of the strong interaction sector represented by the Skyrme and sigma model approach in the 1960-s, before the discovery of QCD. The present electroweak sector of the standard model might represent a "macroscopic" description of electroweak interactions, to be founded microscopically by technicolor like models, in analogy to the aposteriori justification of the Skyrme approach by QCD. Furthermore, the quark-lepton symmetry suggests that quarks too share the fate of leptons and are "macroscopic" objects, made of techniquarks. One might ask *where* do we stop? Unfortunately, as long as the marriage between gravitation and quantum field theory has not been achieved there is no convincing answer to this question and from a philosophical and historical point of view the question *whether* this process stops seems at least as appropriate.

The Skyrme model like any effective theory is of course not renormalisable. Therefore one can expect from it only qualitative results. So are, however, also the conclusions drawn and in particular the relation between spin and quantum numbers, for which, so far, no other explanation has been found.

The identification of all fermions with skyrmions is based on the assumption of the existence of a non-linearly interacting scalar field both in the strong and electro-weak sector. This assumption explains the *universality* of the spin-statistics-quantum number connection and reflects the fact that this relation is a consequence of the identical form of the Lagrangians. That the Pauli principle can be considered as a characteristic property of the Lagrangian is not surprising once one recalls that the class of Lagrangians from which the spin-statistics theorem is derived assumes Lorentz symmetry and space-time commutativity.

Finally, a few words about the "miracles" which some people expect from SUSY in solving the outstanding problems quoted at the beginning. Gauge coupling unification, which addresses the unification of electroweak and strong interactions, however attractive it may look, does not provide the ultimate unification of all interactions, since it leaves gravitation untouched. That dark matter as due to a yet undiscovered s-particle is not much more than the expression of our ignorance about its true cause, since it shifts an unsolved problem from one obscure domain into another one. A similar criticism can be invoked with respect to the relation between SUSY and the hierarchy problem. Ignoring a possible breakdown of the Pauli principle and assuming that supersymmetry may provide the necessary counter-terms for the compensation of the power-law divergences of the radiative corrections to the Higgs mass, its minimal version is plagued by the unsolved mu problem [2], in line with the negative experimental results mentioned in ref. [3]. As a matter of fact, a new approach under the name of *Split Supersymmetry* [26] abandons the invocation of SUSY for the hierarchy problem altogether.

In conclusion the postulate that all fermions are topological solitons may explain the experimentally observed extended connection between spin, statistics and quantum numbers, and the absence of supersymmetry at present energies. Restoration of supersymmetry at higher energies might be associated with the violation of the Pauli exclusion principle.

Acknowledgements**:** I am indebted to David Finkelstein, John Iliopoulos, Tom Kibble, Ady Mann, Slava Rychkov, Adam Schwimmer and Geoffrey Sewell for instructive comments.

**References**

[1] Duck I., Sudarshan, E.C.G.: *Pauli and the Spin-Statistics Theorem,* World Scientific, Singapore (1997).

[2] Cf. e.g. S. P. Martin, *A Supersymmetric Primer*, http://lanl.arxiv.org/pdf/hep-ph/9709356v6.pdf.

[3] P. Bechtle et al. Constrained Supersymmetry after two years of LHC data, http://lanl.arxiv.org/abs/1204.4199.

[4] A. Connes: *Noncommutative Geometry*, Academic Press, 1994; S. Doplicher, K. Fredenhagen, J. E. Roberts, Phys. Lett. B **331**, 39 (1994).

[5] D. Finkelstein and J. Rubinstein, J. Math. Phys. **9,** 1762 (1968).

[6] T. H. R. Skyrme; Proc. Roy. Soc. **260**, 127 (1961); Proc. Roy. Soc. **262**, 237 (1961); J. K. Perring and T. H. R. Skyrme; Nucl. Phys. **31,** 550 (1962); T. H. R. Skyrme; Nucl. Phys. **31,** 556 (1962).

[7] Cf. e.g. R. K. Bhaduri, Models of the Nucleon, *From Quarks to Soliton* (Addison-Wesley 1988; Levant Books 2002, Kolkata).

[8] G. Adkins, C. Nappi, E. Witten, Nucl. Phys. B **228** (552) 1983.

[9] E. Witten, Nucl. Phys. B **223**, 422 (1983).

[10] E. Witten, Nucl. Phys. B **223**, 433 (1983).

[11] G. 't Hooft, Nucl. Phys. B **72,** 461 (1974).

[12] The Skyrme model like any effective theory is of course not renormalisable. Therefore one can expect from it only qualitative results. So are, however, also the conclusions to be discussed below.

[13] T. Nakano et al., Phys. Rev. Lett. **91,** 012002 (2003).

[14] R. M. Weiner, Int. J. Theor. Phys. **49,**1174 (2010).

[15] Cf. T. Appelquist and C. Bernard, Phys. Rev. D **22,** 200 (1980) and literature quoted there.

[16] J. M. Gipson and H. C. Tze, Nucl. Phys. B **183**, 524 (1981), J. M. Gipson, Nucl. Phys. B **231,** 365 (1984).

[17] Tie-zhong Li, Int. J. Theor. Phys. **27,** 307 (1988); for particular realizations of this model cf. Tie-zhong Li, Int. J. Theor. Phys. **30,** 829 (1990).

[18] G. S. Adkins and C. R. Nappi, Nucl. Phys.B **233,** 109 (1984).

[19] The possibility that the Higgs field is represented by a triplet is not only compatible with present data but has also some heuristic advantages as compared with the minimal standard model. Actually, triplet Higgs have become quite popular through the Little Higgs models (cf. e.g. M. Schmaltz and D. Tucker-Smith, *"Little Higgs Review"*, Ann.Rev.Nucl.Part.Sci. **55**, 229, 2005; http://lanl.arxiv.org/abs/hep-ph/0502182). The most recent LHC evidence for a Higgs boson with mass

of the order of 125 GeV supports the "little Higgs" scenario. The considerations of the present paper may be viewed as independent support in this direction, among others also because for little Higgs models there exists the Wess-Zumino term, which, as shown by Witten (Nucl. Phys. B **223** (1983) 422), is responsible for conserved quantum numbers. For this reason one might consider ref. [9] as the basis of the approach to topological interactions in general (cf. R. J. Hill, http://lanl.arxiv.org/abs/0710.5791).

[20] J. Pati and A. Salam, Phys. Rev. D **8** (1973) 1240; R. Foot and H. Lew, Phys.Rev. D **41** (1990) 3502.

[21] Bosonic leptons (baryons) are elementary bosons with lepton (baryon) quantum numbers.

[22] As to the size of the soliton, it is of the order of $(ef)^{-1}$ for $\mathcal{L}_{\text{Skyrme}}$ [7] i.e. $(EF)^{-1}$ for $\mathcal{L}_{\text{Higgs}}$. Since EF is the cut-off where the electroweak symmetry breaks down, the present experimental limits on lepton and quark compositeness would suggest that this cut-off is beyond 1 TeV, in agreement with Litle Higgs models.

[23] S. Chandrasekhar, Astrophys. J. **74**, 81 (1931); Rev. Mod. Phys. **56**, 137 (1984).

[24] M. Chaichian, K. Nishijima and A. Tureanu, Phys. Lett. B **568,** 146 (2003).

[25] A. Tureanu, Phys.Lett. B638 (2006) 296-301, hep-th/0603219.

[26] N. Arkani-Hamed and S. Dimpoulos, JHEP **0506** 073 (2005) [arXiv:hep-th/0405159]; G. F. Giudice and A. Romanino, Nucl. Phys. B **699** (2004) 65 [arXiv:hep-ph/0406088]; N. Arkani-Hamed, S. Dimopoulos, G. F. Giudice and A. Romanino, Nucl. Phys. B **709** (2005) 3 [arXiv:hep-ph/0409232].

17